\documentclass[a4paper]{jpconf}
\usepackage{graphicx}
\usepackage{amsmath}
\usepackage{citesort}
\begin{document}
\title{From nucleons to nuclei to fusion reactions}

\author{S. Quaglioni$^1$, P. Navr\'atil$^{2,1}$, R. Roth$^3$, and W. Horiuchi$^4$}

\address{$^1$ Lawrence Livermore National Laboratory, P.O. Box 808, L-414, Livermore, CA 94551, USA}
\address{$^2$ TRIUMF, 4004 Wesbrook Mall, Vancouver, BC V6T 2A3, Canada}
\address{$^3$ Institut f\"{u}r Kernphysik, Technische Universit\"{a}t Darmstadt, 64289 Darmstadt, Germany}
\address{$^4$ RIKEN Nishina Center, Wako 351-0198, Japan}
\ead{quaglioni1@llnl.gov}

\begin{abstract}
Nuclei are prototypes of many-body open quantum systems.
Complex aggregates of protons and neutrons that interact through forces arising from quantum chromo-dynamics, nuclei exhibit both bound and unbound states, which can be strongly coupled. In this respect, one of the major challenges for computational nuclear physics, is to provide a unified description of structural and reaction properties of nuclei that is based on the fundamental underlying physics: the constituent nucleons and the realistic interactions among them. This requires a combination of innovative theoretical approaches and high-performance computing. In this contribution, we present one of such promising techniques, the {\em ab initio} no-core shell model/resonating-group method, and discuss applications to light nuclei scattering and fusion reactions that power stars and Earth-base fusion facilities. 
\end{abstract}

\section{Introduction}
Low-energy nuclear reactions are the fuel of stars such as our Sun, but also of research facilities directed toward developing fusion power by either magnetic ({\em e.g.}\ ITER~\cite{ITER}) or inertial ({\em e.g.}\ NIF~\cite{NIF}) confinement.  Consequently, astrophysics models as well as fusion-experiment simulations rely on various nuclear input data such as cross sections (or, equivalently, astrophysical S-factors), energy spectra, angular distributions, {\em etc.}, for thermonuclear reactions.

Providing the research community with accurate nuclear fusion data is one of the longstanding challenges of experimental and theoretical nuclear physics both.  On one hand, due to the extreme low energies and ionized state of matter at which they take place, nuclear fusion reactions can be very challenging or even impossible to measure in beam-target experiments, leaving to theory a large role in extracting (often extrapolating) the astrophysically important information. On the other hand, a fully developed fundamental theory able to provide accurate predictions and uncertainties for a large range of relevant reactions is still missing, the main obstacle being the treatment of scattering states for many-nucleon systems based on the constituent nucleons and the realistic interactions among them. 

Some of the outstanding light-nucleus uncertainty sources in astrophysics applications include: reactions leading to the nucleosynthesis of $^8$B (and the production of the solar neutrinos measured in terrestrial experiments) such as 
the $^7$Be$(p,\gamma)^8$B and $^3$He$(\alpha,\gamma)^7$Be radiative capture rates;  the thermonuclear reaction rates of $\alpha$ capture on $^8$Be and $^{12}$C nuclei during the stellar helium burning; and fusion reactions that affect the predictions of Big Bang nucleosyntesys for the abundances of light elements, such as the $^3$He($d$,$p$)$^4$He. At the same time, large uncertainties in reactions such as the $^3$H$+d \rightarrow ^4$He$+n+\gamma$ bremsstrahlung process or the $^3$H$(^3$H$,2n)^4$He fusion are limiting factors in the understanding of how the fuel is assembled in deuterium-tritium based fusion experiments. The increasingly sophisticate experimental effort devoted to these and other reactions needs to be accompanied by accurate calculations with predictive capability in the low-energy limit.

At the same time, developing a comprehensive description of nuclear properties requires also understanding exotic nuclei, short-lived nuclei that inhabit remote regions of the nuclear landscape, where the neutron-to-proton rations are larger or smaller than those naturally occurring on earth. These nuclei are difficult to study, due to their fragile nature and the often small production cross section. They present new phenomena, such as halo densities, vanishing of magic numbers, and ``abnormal" spin-parity of ground states. Once again, the new experiments with exotic nuclei that will be carried out at the Facility for Rare Isotope Beams need to be supported by predictive calculations.

In both instances, a fundamental theory is needed that will enhance our ability of predicting nuclear behaviors in regions that might not be accessible to experiments. Developing such a theory requires abandoning the ``traditional'' separated treatment of discrete states and scattering continuum, and seeking a unified description of structural and reaction properties. 
Indeed, nuclei are prototypical many-body open quantum systems. They exhibit bound states, resonances, scattering states, all of which can be strongly coupled. Nuclear structure properties in the vicinity of thresholds are affected by the continuum of scattering and decay channels. At the same time, low-energy scattering and reactions, such as fusions rates, are affected by the internal structure of the interacting nuclei. 

In the framework of the interacting shell model, first attempts to incorporate continuum effects in structure calculations 
 based on the Feshbach projection formalism~\cite{Herman1958357,Herman1962287} 
led to various formulations of the continuum shell model~\cite{MW1969,Barz1977111,Rotter1978237,BNOP1999,PhysRevC.74.064314}. 
Modern versions of this are the shell model embedded in the continuum~\cite{BNOP1999,Rotureau200613}, and the time-dependent approach to the continuum shell model~\cite{PhysRevC.79.044308}. A more recent attempt of extending the interacting shell model to the treatment of open quantum systems is the complex-energy or Gamow shell model, in which a symmetric description of bound, resonant and scattering single-particle states is achieved by working within a Berggren ensemble~\cite{PhysRevLett.89.042502,PhysRevLett.89.042501,PhysRevC.67.054311,PhysRevC.84.051304}.  The cluster orbital shell model is an other variant of complex-energy shell model~\cite{PhysRevC.75.034316}. 

An alternative to the shell model picture of the nucleus was proposed in 1937 by Wheeler~\cite{PhysRev.52.1083}, in the form of the resonating-group method~\cite{Tang1978167}, a microscopic approach which explicitly takes cluster correlations into account. Here, nuclear bound states, resonances and reactions are treated within the same framework by means of expansions over fully antisymmetric cluster wave functions.  A modern and physically equivalent incarnation of this approach is the generator coordinate method~\cite{Horiouchi,PhysRevC.70.065802,PhysRevC.72.024309,PhysRevC.80.044310}. In this techniques, which make often use of central effective $NN$ interactions and simplified cluster wave functions that are not necessarily eigenstates of the chosen Hamiltonian, the clustering of nuclei is introduced explicitly, so that the treatment of reactions becomes straightforward.

Finally, a unified description of the structural and reaction properties of light-to-medium mass nuclei is recently starting to be accomplished also within an {\em ab initio} framework, thanks mainly to the Green's function Monte Carlo~\cite{PhysRevLett.99.022502}, no-core shell model/resonating-group method (NCSM/RGM)~\cite{PhysRevLett.101.092501,PhysRevC.79.044606,PhysRevC.83.044609} and coupled cluster technique with a Gamow-Hartree-Fock basis~\cite{PhysRevLett.104.182501}. 

In this contribution, we give an overview of the NCSM/RGM approach. In Sec.~\ref{formalism} we briefly present the NCSM/RGM formalism, while a summary of its most important applications is given in Sec.~\ref{applications}. Conclusions are drawn in Sec.~\ref{conclusions}.

\section{{\em Ab initio} NCSM/RGM}
\label{formalism}
The {\em ab initio} nuclear reaction approach that we are developing is an extension of the {\em ab initio} no-core shell model (NCSM)~\cite{PhysRevLett.84.5728}. The innovation which allows us to go beyond bound states and treat reactions is the use of cluster basis states in the spirit of the resonating-group method,
\begin{equation}
|\Phi^{J^\pi T}_{\nu r}\rangle = \Big [ \big ( \left|A{-}a\, \alpha_1 I_1^{\,\pi_1} T_1\right\rangle \left |a\,\alpha_2 I_2^{\,\pi_2} T_2\right\rangle\big ) ^{(s T)}
\,Y_{\ell}\left(\hat r_{A-a,a}\right)\Big ]^{(J^\pi T)}\,\frac{\delta(r-r_{A-a,a})}{rr_{A-a,a}}\,,\label{basis}
\end{equation}
in which each nucleon cluster is described within the NCSM. The above translational invariant cluster basis states describe two nuclei (a target and a projectile composed of $A-a$ and $a$ nucleons, respectively) whose centers of mass are separated by the relative coordinate $\vec r_{A-a,a}$ and that are traveling in a $^{2s}\ell_J$ wave or relative motion (with $s$ the channel spin, $\ell$ the relative momentum, and $J$ the total angular momentum of the system). Additional quantum numbers characterizing the basis states are  parity $\pi=\pi_1\pi_2(-1)^{\ell}$ and total isospin $T$. For the intrinsic (antisymmetric) wave functions of the two nuclei we employ the eigenstates $\left|A{-}a\, \alpha_1 I_1^{\,\pi_1} T_1\right\rangle$ and $\left |a\,\alpha_2 I_2^{\,\pi_2} T_2\right\rangle$ of the $(A-a)$- and $a$-nucleon intrinsic Hamiltonians, respectively, as obtained within the NCSM approach.  These are characterized by the spin-parity, isospin and energy labels $I_i^{\pi_i},T_i$, and $\alpha_i$, respectively, where $i=1,2$.  In our notation, all these quantum numbers are grouped into a cumulative index $\nu=\{A{-}a\,\alpha_1I_1^{\,\pi_1} T_1;\, a\, \alpha_2 I_2^{\,\pi_2} T_2;\, s\ell\}$. Finally, we note that the channel states~(\ref{basis}) are not antisymmetric with respect to exchanges of nucleons pertaining to different clusters. Therefore, to preserve the Pauli principle one has to introduce the appropriate inter-cluster antisymmetrizer, schematically
\begin{equation}
\hat{\mathcal A}_{\nu}=\sqrt{\frac{(A{-}a)!a!}{A!}}\left( 1+\sum_{P\neq id}(-)^pP\right)\,,
\label{antisymmetrizer}
\end{equation}   
where the sum runs over all possible permutations of nucleons $P$ different from the identical one 
that can be carried out between two different clusters (of $A-a$ and $a$ nucleons, respectively), and $p$ is the number of interchanges characterizing them. The operator~(\ref{antisymmetrizer}) is labeled by the channel index $\nu$ to signify that its form depends on the mass partition, $(A-a,a)$, of the channel state to which is applied. 

The channel states~(\ref{basis}), fully antisimmetrized by the action of the antisymmetrization operator $\hat A_\nu$, are used as a continuous basis set to expand the many-body wave function,   
\begin{eqnarray}
|\Psi^{J^\pi T}\rangle &=& \sum_{\nu} \int dr r^2 \, \hat{\mathcal A}_{\nu}|\Phi^{J^\pi T}_{\nu r}\rangle  \frac{[{\cal N}^{-1/2}\chi]^{J^\pi T}_\nu(r)}{r}
\, , \label{trial}
\end{eqnarray}
where $\chi^{J^\pi T}_\nu(r)$ represent continuous linear variational amplitudes that are determined by solving the orthogonalized RGM equations: 
\begin{equation}
{\sum_{\nu^\prime}\int dr^\prime r^{\prime\,2}} [{\mathcal N}^{-\frac12}{\mathcal H}\,{\mathcal N}^{-\frac12}]^{J^\pi T}_{\nu\nu^\prime\,}(r,r^\prime)\frac{\chi^{J^\pi T}_{\nu^\prime} (r^\prime)}{r^\prime} = E\,\frac{\chi^{J^\pi T}_{\nu} (r)}{r}  \label{RGMeq}.
\end{equation}
Here ${\mathcal N}^{J^\pi T}_{\nu\nu^\prime}(r, r^\prime)$ and ${\mathcal H}^{J^\pi T}_{\nu\nu^\prime}(r, r^\prime)$, commonly referred to as integration kernels, are respectively the overlap (or norm) and Hamiltonian matrix elements over the antisymmetrized basis~(\ref{basis}), {\em i.e.}:  
\begin{align}
{\mathcal N}^{J^\pi T}_{\nu^\prime\nu}(r^\prime, r) = \left\langle\Phi^{J^\pi T}_{\nu^\prime r^\prime}\right|\hat{\mathcal A}_{\nu^\prime}\hat{\mathcal A}_{\nu}\left|\Phi^{J^\pi T}_{\nu r}\right\rangle\,, 
&\qquad
{\mathcal H}^{J^\pi T}_{\nu^\prime\nu}(r^\prime, r) = \left\langle\Phi^{J^\pi T}_{\nu^\prime r^\prime}\right|\hat{\mathcal A}_{\nu^\prime}H\hat{\mathcal A}_{\nu}\left|\Phi^{J^\pi T}_{\nu r}\right\rangle
\label{NH-kernel}
\end{align}
%
%
%
%
%
where $H$ is the microscopic $A-$nucleon Hamiltonian and $E$ is the total energy in the center of mass (c.m.)\ frame.
The calculation of the above many-body matrix elements, which contain all the nuclear structure and antisymmetrization properties of the system under consideration, represents the main task in performing RGM calculations. In the following we will review the various steps required for one of such calculations within the NCSM/RGM approach.

\subsection{Input: nuclear Hamiltonian and cluster eigenstates} 
We start from the microscopic Hamiltonian for the $A-$nucleon system,
\begin{equation}
H=\frac{1}{A}\sum_{i<j=11}^A\frac{(\vec{p}_i-\vec{p}_j)^2}{2m}+\sum_{i<j=1}^A V^{NN}_{ij}+\sum_{i<j<k=1}^A V^{NNN}_{ijk},
\label{H}
\end{equation}	
where $m$ is the nucleon mass, and $V^{NN}$ and $V^{NNN}$ the nucleon-nucleon ($NN$) --nuclear plus point-Coulomb-- and three-nucleon ($NNN$) interactions, respectively. For the purpose of the RGM approach, it is convenient to separate Eq.~(\ref{H}) into the intrinsic Hamiltonians for the $(A-a)$- and $a$-nucleon systems, respectively $H_{(A-a)}$ and $H_{(a)}$, plus the relative motion Hamiltonian according to:
\begin{equation}\label{Hamiltonian}
H=T_{\rm rel}(r)+\bar{V}_{\rm C}(r)+{\mathcal V}_{\rm rel} +H_{(A-a)}+H_{(a)}\,.
\end{equation}
Here, $T_{\rm rel}(r)$ is the relative kinetic energy, $\bar{V}_{\rm C}(r)=Z_{1\nu}Z_{2\nu}e^2/r$ ($Z_{1\nu}$ and $Z_{2\nu}$ being the charge numbers of the clusters in channel $\nu$) the average Coulomb interaction between pairs of clusters, and ${\mathcal V}_{\rm rel}$ is localized relative (inter-cluster) potential given by:
\begin{eqnarray}
{\mathcal V}_{\rm rel} &=& \sum_{i=1}^{A-a}\sum_{j=A-a+1}^AV^{NN}_{ij} + \sum_{i<j=1}^{A-a}\sum_{k=A-a+1}^AV^{NNN}_{ijk} + \sum_{i=1}^{A-a}\sum_{j<k=A-a+1}^AV^{NNN}_{ijk} - \bar{V}_{\rm C}(r)\label{pot}\,.
\end{eqnarray}
Besides the nuclear components of the interactions between nucleons belonging to different clusters, it is important to notice that the overall contribution to the relative potential~(\ref{pot}) coming from the Coulomb interaction,
\begin{equation}
\sum_{i=1}^{A-a}\sum_{j=A-a+1}^A\left(\frac{e^2(1+\tau^z_i)(1+\tau^z_j)}{4|\vec r_i-\vec r_j|} -\frac{1}{(A-a)a}\bar V_{\rm C}(r)\right)\,,
\end{equation}
 is also localized, presenting an $r^{-2}$ behavior, as the distance $r$ between the two clusters increases.
 
The other main input required for calculating the RGM integration kernels of Eq.~(\ref{NH-kernel}) are the eigenstates of the projectile and target wave functions. In the NCSM/RGM approach, these are obtained  by diagonalizing $H_{(A-a)}$ and $H_{(a)}$ in the model spaces spanned by the $(A-a)$- and $a$-nucleon NCSM bases, respectively. We adopt complete HO bases, the size of which is defined by the maximum number, $N_{\rm max}$, of HO quanta above the lowest configuration shared by the nucleons  (the definition of the model-space size coincides for eigenstates of the same parity, differs by one unity for eigenstates of opposite parity). The same $N_{\rm max}$ value and HO frequency $\Omega$ are used for both clusters.
Thanks to the unique properties of the HO basis,  we can make use of Jacobi-coordinate wave functions~\cite{three_NCSM,Jacobi_NCSM} for both nuclei or only for the lightest of the pair (typically the projectile, $a\le4$), and still preserve the translational invariance of the problem. In the second case we expand the eigenstates of the heavier cluster (typically the target) on a Slater-determinant (SD) basis, and remove completely the spurious c.m.\ components as explained in Sec. II.B.2 of Ref.~\cite{PhysRevC.79.044606}. Such dual approach can be used as a way of verifying our results.  The use of the SD basis is computationally advantageous and allows us to explore reactions involving $p$-shell nuclei.  In this case, the Hamiltonian matrix can reach dimensions up to $10^9$, and diagonalizations are obtained by means of the Lanczos algorithm~\cite{Lanczos,Wilkinson}.

Because of the complexity of the nuclear force among protons and neutrons, most nuclear interaction models generate strong short-range nucleon-nucleon correlations and the large but finite model spaces computationally achievable are not sufficient to reach the full convergence through a ``bare'' calculation. In these cases it is crucial to make the nuclear many-body problem computationally more tractable by means of effective interactions obtained through unitary transformations of the initial Hamiltonian. Such effective interactions can be derived, for target, projectile and compound $A$-nucleon system, through the Lee-Suzuki procedure in a consistent and formally exact way within the NCSM basis~(see Refs.~\cite{PhysRevLett.84.5728,PhysRevC.62.054311} and Sec. II.B of Ref.~\cite{PhysRevC.79.044606}).  Alternatively, one can perform variational calculations using bare similarity-renormalization-group (SRG)~\cite{PhysRevC.75.061001,PhysRevC.77.064003} evolved potentials. This second choice is preferable in the NCSM/RGM approach, where, to avoid possible inconsistencies, it is desirable that the same nuclear interaction be used to obtain the structure of projectile and target, as well as the overall projectile-target potential ({\em i.e.}, in $H_{(A-a)}$, $H_{(a)}$, and ${\mathcal V}_{\rm rel}$).  In most of the applications presented in Sec.~\ref{applications} we employ SRG-evolved chiral N$^3$LO~\cite{N3LO} $NN$ potentials (SRG-N$^3$LO).

\subsection{Calculation of norm and Hamiltonian kernels}
\label{kernels}
From Eqs.~(\ref{antisymmetrizer}) and~(\ref{Hamiltonian}) it follows that the norm and Hamiltonian kernels can be factorized into ``full-space'' and ``model-space" components according to:
\begin{align}
{\mathcal N}^{J^\pi T}_{\nu^\prime\nu}(r^\prime, r)
&= \delta_{\nu^\prime\nu}\frac{\delta(r^\prime-r)}{r^\prime r} + {\mathcal N}^{\rm ex}_{\nu^\prime\nu}(r^\prime, r)
\label{N-kernel-2}
\end{align}
and,
\begin{align}
{\mathcal H}^{J^\pi T}_{\nu^\prime\nu}(r^\prime, r) &= \left[{T}_{\rm rel}(r')+\bar{V}_C(r')+E_{\alpha_1'}^{I_1'T_1'} +E_{\alpha_2'}^{I_2'T_2'}\right]
\mathcal{N}_{\nu'\nu}^{J^\pi T}(r', r)+\mathcal{V}^{J^\pi T}_{\nu' \nu}(r',r)\,,
\label{H-kernel-2}
\end {align}
where the exchange part of the norm, ${\mathcal N}^{\rm ex}_{\nu^\prime\nu}(r^\prime, r)$, and the potential kernel, $\mathcal{V}^{J^\pi T}_{\nu' \nu}(r',r)$, are obtained in the truncated model space by expanding the  Dirac delta function of Eq.~(\ref{basis}) on a set of HO radial wave functions with identical frequency $\Omega$ and model-space size $N_{\rm max}$ as those used for the two clusters, more in detail:
\begin{align}
{\mathcal N}^{\rm ex}_{\nu^\prime\nu}(r^\prime, r) &= 
\sum_{n^\prime n}R_{n^\prime\ell^\prime}(r^\prime)R_{n\ell}(r) 
\times \left\{
\begin{array}{ll}
	\left\langle\Phi^{J^\pi T}_{\nu^\prime n^\prime}\right| \sum_{P\neq id}(-)^p P \left|\Phi^{J^\pi T}_{\nu n}\right\rangle& \quad {\rm if}~a^\prime=a\\
	\\
	\left\langle\Phi^{J^\pi T}_{\nu^\prime n^\prime}\right| \sqrt{\tfrac{ A! }{ (A-a^\prime)! a^\prime !}} \hat {\mathcal A}_\nu \left|\Phi^{J^\pi T}_{\nu n}\right\rangle& \quad {\rm if}~a^\prime\neq a
\end{array}
\right .
\end{align}
and 
\begin{align}
\mathcal{V}^{J^\pi T}_{\nu' \nu}(r',r) &= 
	\sum_{n^\prime n}R_{n^\prime\ell^\prime}(r^\prime)R_{n\ell}(r)\left\langle\Phi^{J^\pi T}_{\nu^\prime r^\prime}\right| 
	\sqrt{\tfrac{ A!}{ (A-a^\prime)! a^\prime !}} {\mathcal V}_{\rm rel}\hat{\mathcal A}_{\nu} 
	\left|\Phi^{J^\pi T}_{\nu r}\right\rangle\,.
\end{align}
Such a procedure is justified for matrix elements of localized operators such as those entering the exchange part of the norm and the potential kernels. To obtain the above expressions, we introduce the HO Jacobi channel states
\begin{equation}
|\Phi^{J^\pi T}_{\nu n}\rangle = \Big [ \big ( \left|A{-}a\, \alpha_1 I_1^{\,\pi_1} T_1\right\rangle \left |a\,\alpha_2 I_2^{\,\pi_2} T_2\right\rangle\big ) ^{(s T)}\,Y_{\ell}\left(\hat \eta_{A-a}\right)\Big ]^{(J^\pi T)}\,R_{n\ell}(r_{A-a,a})\,,\label{ho-basis-n}
\end{equation}
take advantage of the commutation between antisymmetrizers~(\ref{antisymmetrizer}) and $A$-nucleon Hamiltonian~(\ref{H}), $[\hat{\mathcal A}_{\nu},H]{=}0$, and use the following relationship dictated by symmetry considerations:
\begin{align}
\hat{\mathcal A}_{\nu^\prime}\hat{\mathcal A}_{\nu}|\Phi^{J^\pi T}_{\nu n}\rangle =  \sqrt{\tfrac{ A!}{ (A-a^\prime)! a^\prime !}} \hat{\mathcal A}_{\nu}|\Phi^{J^\pi T}_{\nu n}\rangle\,. 
\end{align}
As pointed out in Sec.~\ref{formalism}, the channel states~(\ref{basis}) are not anti-symmetric with respect to the exchange of nucleons pertaining to different clusters (fully anti-symmetric states are recovered  through the action of the operator $\hat{\mathcal A}_{\nu}$). As a consequence, the Hamiltonian kernel as defined in Eq.~(\ref{H-kernel-2}) is explicitly non Hermitian. Using $\hat{\mathcal A}_{\nu^\prime}H\hat{\mathcal A}_{\nu}=\frac12(\hat{\mathcal A}_{\nu^\prime}\hat{\mathcal A}_{\nu}H+H\hat{\mathcal A}_{\nu^\prime}\hat{\mathcal A}_{\nu})$, we introduce the Hermitized Hamiltonian kernel $\bar{\mathcal H}^{J^\pi T}_{\nu^\prime\nu}$ in the form
\begin{equation}
\bar{\mathcal H}^{J^\pi T}_{\nu^\prime\nu}(r^\prime,r)\!=\!\left\langle\Phi^{J^\pi T}_{\nu^\prime r^\prime}\right| \tfrac 12 \left(\hat{\mathcal A}_{\nu^\prime} H  \sqrt{\tfrac{ A! }{ (A-a)! a!}} + 
	\sqrt{\tfrac{ A!}{ (A-a^\prime)! a^\prime !}} H \hat{\mathcal A}_{\nu} 
	 \right)\left| \Phi^{J^\pi T}_{\nu r}\right\rangle.\label{ham-herm}
\end{equation}
As a final note, we would like to point out that the exchange part of the norm kernel is explicitly hermitian, in particular:
\begin{equation}
	\left\langle\Phi^{J^\pi T}_{\nu^\prime n^\prime}\right| \sqrt{\tfrac{ A! }{ (A-a^\prime)! a^\prime !}} \hat {\mathcal A}_\nu \left|\Phi^{J^\pi T}_{\nu n}\right\rangle = \left\langle\Phi^{J^\pi T}_{\nu^\prime n^\prime}\right| \hat {\mathcal A}_{\nu^\prime} \sqrt{\tfrac{ A! }{ (A-a)! a!}}  \left|\Phi^{J^\pi T}_{\nu n}\right\rangle\,.
\end{equation}

\begin{figure}[t]
\begin{center}
\includegraphics*[width=0.4\columnwidth]{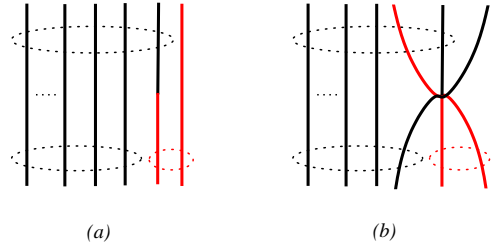}
\end{center}
\caption{Diagrammatic representation of: ($a$) ``overlap" and ($b$) ``one-nucleon-exchange" components of the norm kernel of Eq.~(\ref{norm-ex-coupled}). The groups of circled lines represent the $(A{-}2)$-, $(A{-}1)$-, and two-nucleon clusters. Bottom and upper part of the diagram represent initial and final states, respectively.}\label{diagram-norm}
\end{figure}
The explicit form of the inter-cluster antisymmetrizers for $a=1$ and $a=2$ projectiles, together with the algebraic expressions of the integration kernels for the specific cases $a^\prime=a=1$ and $a^\prime=a=2$ of equal mass partitions in initial and final states can be found in Refs.~\cite{PhysRevC.79.044606} and~\cite{PhysRevC.83.044609}, respectively.
For reactions involving a deuterium-nucleus entrance and nucleon-nucleus exit channels [{\em e.g.}, $^3$H$(d,n)^4$He] or vice versa, and, more in general, whenever both nucleon-nucleus and deuterium-nucleus channel basis states are used in the RGM model space, one has to address the additional contributions coming from the off-diagonal matrix elements between the two mass partitions: $(A-1,1)$ and $(A-2,2)$. Here, we list these additional terms (two for the norm and five for the Hamiltonian kernel), without entering in the details of their algebraic expressions, which will be published elsewhere.

 The exchange part of the norm kernel for an $(A-2,2)$ mass partition in the initial state ($\nu$) and an $(A-1,1)$ mass partition in the final state ($\nu^\prime$) can be cast in form:
\begin{align}
	{\mathcal N}^{\rm ex}_{\nu^\prime\nu}(r^\prime, r) = &
	 \sqrt{\tfrac{A-1}2} \sum_{n^\prime\,n} R_{n^\prime\ell^\prime}(r^\prime) R_{n\ell}(r) \left[ 2 \left\langle\Phi^{J^\pi T}_{\nu^\prime n^\prime}\right|\left.\Phi^{J^\pi T}_{\nu n}\right\rangle  - (A-2) \left\langle\Phi^{J^\pi T}_{\nu^\prime n^\prime}\right|\hat P_{A-2,A}\left|\Phi^{J^\pi T}_{\nu n}\right\rangle \right]\,.
	 \label{norm-ex-coupled}
\end{align} 
Two terms contribute to the above equation: the overlap between the initial $(A-2,2)$ and final (A-1,1) binary-cluster states [corresponding to diagram $(a)$ of Fig.~\ref{diagram-norm}]; and a one-nucleon exchange term, corresponding to diagram $(b)$ of  Fig.~\ref{diagram-norm}. The corresponding Hermitized Hamiltonian kernel is obtained according to Eq.~(\ref{ham-herm}). In particular, the $NN$ part of the Hermitized potential kernel is given by: 
 \begin{align}
	\bar {\mathcal V}^{NN}_{\nu^\prime\nu}(r^\prime, r) = &
	 \sqrt{\tfrac{A-1}2} \sum_{n^\prime\,n} R_{n^\prime\ell^\prime}(r^\prime) R_{n\ell}(r) \left[ 2(A-2) \left\langle\Phi^{J^\pi T}_{\nu^\prime n^\prime}\right| V_{A-2,A}(1-\hat P_{A-2,A})\left|\Phi^{J^\pi T}_{\nu n}\right\rangle \right. \nonumber \\[2mm]
	&\quad  + \left\langle\Phi^{J^\pi T}_{\nu^\prime n^\prime}\right| V_{A-1,A}\left|\Phi^{J^\pi T}_{\nu n}\right\rangle 
	+ (A-2) \left\langle\Phi^{J^\pi T}_{\nu^\prime n^\prime}\right| V_{A-2,A-1}\left|\Phi^{J^\pi T}_{\nu n}\right\rangle \nonumber\\[2mm]
	 &\quad \left. -(A-2)(A-3) \left\langle\Phi^{J^\pi T}_{\nu^\prime n^\prime}\right| \tfrac 12 \hat P_{A-2,A}V_{A-3,A-2} + V_{A-3,A-2} \hat P_{A-2,A} \left|\Phi^{J^\pi T}_{\nu n}\right\rangle \right ] \,.
	 \label{pot-coupled}
\end{align} 
\begin{figure}[t]
\begin{center}
\includegraphics*[width=0.98\columnwidth]{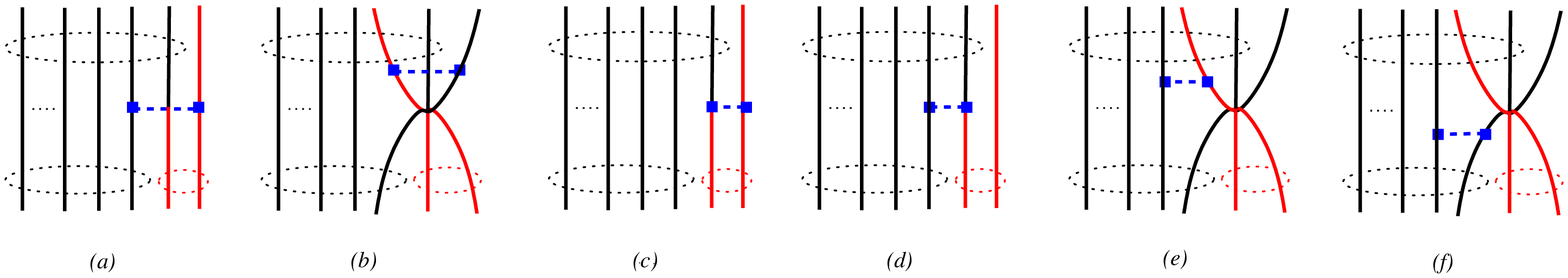}
\end{center}
\caption{Diagrammatic representation of the components of the Hermitized potential kernel. The groups of circled lines represent the $(A{-}2)$-, $(A{-}1)$-, and two-nucleon clusters. Bottom and upper part of the diagram represent initial and final states, respectively.}\label{diagram-pot}
\end{figure}
In this expression, we identify four separate terms corresponding to the six diagrams of Fig.~\ref{diagram-pot}. The first term on the right-hand-side of Eq.~(\ref{pot-coupled}) corresponds to diagrams $(a)$ and $(b)$, the second and third terms correspond to diagrams $(c)$ and $(d)$, respectively, while diagrams $(e)$ and $(f)$ represent the fourth term.

The applications presented in this contribution are obtained with the $NN$ part only of the inter-cluster interaction. The inclusion of the three-nucleon force into the formalism, although more involved, is straightforward and is currently in progress. As an example, the potential kernel for the same $(A-1,1)$ partition in both initial and final states ($a^\prime=a=1$) contains two additional terms due to the presence of the $NNN$ force in the Hamiltonian. These are:
 \begin{align}
	{\mathcal V}^{NNN}_{\nu^\prime\nu}(r^\prime, r) = &
	 \sum_{n^\prime\,n} R_{n^\prime\ell^\prime}(r^\prime) R_{n\ell}(r) \left[ \tfrac{(A-1)(A-2)}{2} \left\langle\Phi^{J^\pi T}_{\nu^\prime n^\prime}\right| V_{A-2,A-1,A}(1-2\hat P_{A-1,A})\left|\Phi^{J^\pi T}_{\nu n}\right\rangle \right.\nonumber \\[2mm]
	&\qquad  \left.+\tfrac{(A-1)(A-2)(A-3)}{2} \left\langle\Phi^{J^\pi T}_{\nu^\prime n^\prime}\right| \hat P_{A-1,A}V_{A-3,A-2,A-1}\left|\Phi^{J^\pi T}_{\nu n}\right\rangle 
	 \right ] \,.
	 \label{pot-NNN}
\end{align} 
\begin{figure}[h]
\begin{center}
\includegraphics*[width=0.55\columnwidth]{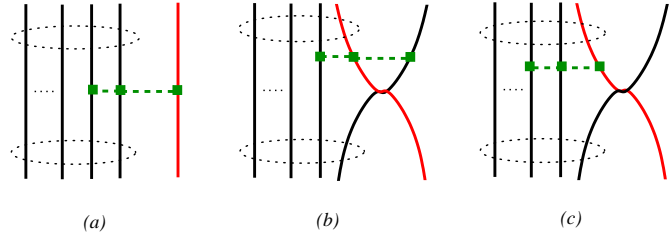}
\end{center}
\caption{Diagrammatic representation of the components of the direct ($(a)$ and $(b)$) and exchange components of the $NNN$ potential kernel for the same $(A-1,1)$ partition in both initial and final states ($a^\prime=a=1$). The groups of circled lines represent the $(A{-}1)$-nucleon cluster. Bottom and upper part of the diagram represent initial and final states, respectively.}\label{diagram-NNNpot}
\end{figure}
As for the corresponding $NN$ portion of the potential kernel, there are a direct and an exchange term, described by diagrams $(a)$ and $(b)$, and diagram $(c)$ of Fig.~\ref{diagram-NNNpot}, respectively.

The calculation of matrix elements of the type shown in Eqs.~(\ref{norm-ex-coupled})-(\ref{pot-NNN}) represents the most computationally intensive step of the NCSM/RGM approach. Each of them requires the derivation and implementation in specialized algorithms of non trivial algebraic expressions.  An advantage of the NCSM/RGM approach is that one can work with Slater-determinant target wave functions and still preserve the translational invariance of the problem (see Sec. II.B.2 of Ref.~\cite{PhysRevC.79.044606}). This allows the use of powerful second quantization techniques. In such a framework, the kernels are written in terms of matrix elements of the one-, two- , three-body densities, {\em etc.}, calculated on the target wave functions. 
As an example,  the first two [diagrams $(a)$ and ($b$) of Fig.~\ref{diagram-NNNpot}] and the last [diagram $(c)$ of Fig.~\ref{diagram-NNNpot}] terms of the $NNN$ projectile kernel~(\ref{pot-NNN}) depend on two- and three-body density matrix elements of the target nucleus, respectively. In comparison, the $NN$ potential kernel for an identical $(A-1,1)$ mass partition in both initial and final states depends only on matrix elements of the one and two-body densities. This highlights the increased complexity of the NCSM/RGM approach when the three-nucleon force is included in the Hamiltonian.  To conclude this section, we note that the computational demand (in terms of both CPU time and memory requirements) for the calculation of the integration kernels rapidly increases with projectile mass, number of projectile/target states, and number of channels included making the NCSM/RGM a computationally intensive approach.

\subsection{Output: eigenstates, eigenenergies and Scattering matrix}
Due to the Pauli exclusion principle, the integration kernels 
are surfaces in three dimensions, and give rise to channel-dependent non-local couplings $W^{J^\pi T}_{\nu \nu^\prime}(r,r^\prime)$ between the unknown projectile-target relative wave functions. Indeed, by separating local diagonal and non-local terms, Eq.~(\ref{RGMeq}) can be cast in the form:
\begin{align}
[\hat T_{\rm rel}(r) + \bar V_{\rm C}(r) -(E - E_{\alpha_1}^{I_1^{\pi_1} T_1} - E_{\alpha_2}^{I_2^{\pi_2} T_2})]\frac{\chi^{J^\pi T}_{\nu} (r)}{r} + \sum_{\nu^\prime}\int dr^\prime\,r^{\prime\,2} \,W^{J^\pi T}_{\nu \nu^\prime}(r,r^\prime)\,\frac{\chi^{J^\pi T}_{\nu^\prime}(r^\prime)}{r^\prime} = 0.\label{r-matrix-eq}
\end{align} 
The solution of such a set of integral-differential coupled channel equations represents a fairly standard problem in scattering theory, only slightly complicated by the presence of non local coupling potentials.  A particularly efficient technique for the solution of Eq.~(\ref{r-matrix-eq}) is the $R$-matrix method on Lagrange mesh~\cite{Hesse199837,Hesse2002184}. This is a method based on the microscopic $R$-matrix theory~\cite{RevModPhys.30.257,R-matrix} in which one assumes that $\bar{V}_{\rm C}(r)$ is the only interaction experienced by the clusters beyond a sufficiently large separation $r_0$, 
thus dividing the configuration space into an internal and an external region. In the internal region, the wave function can be written as an expansion over  a set of square integrable basis functions, 
while in the external region it can be approximated by its asymptotic form for large $r$, 
\begin{equation}
\chi^{J^\pi T}_\nu(r) = \frac{\rm i}{2} v_{\nu}^{-1/2}[\delta_{\nu i}H^{-}_{\ell}(\eta_{\nu},\kappa_{\nu}r)-S^{J^\pi T}_{\nu i} H^{+}_{\ell}(\eta_{\nu},\kappa_{\nu}r)]\,,\label{scattering}
\end{equation}
for scattering states, or
\begin{equation}
\chi^{J^\pi T}_\nu(r) = C^{J^\pi T}_\nu\,W_\ell(\eta_\nu,\kappa_\nu r)\,,\label{bound}
\end{equation}
for bound states.  Here,  
$H^{\mp}_{\ell}(\eta_{\nu},\kappa_{\nu}r)=G_{\ell}(\eta_{\nu},\kappa_{\nu}r)\mp {\rm i} F_{\ell}(\eta_{\nu},\kappa_{\nu}r)$
are incoming and outgoing Coulomb functions, whereas $W_\ell(\eta_\nu,\kappa_\nu r)$ are Whittaker functions. They depend on the channel state relative angular momentum $\ell$,  wave number $\kappa_\nu$, and Sommerfeld parameter $\eta_\nu$. The corresponding velocity is denoted as $v_{\nu}$. 
The scattering matrix $S^{J^\pi T}_{\nu i}$ ($i$ being the initial channel) in Eq.~(\ref{scattering}) is obtained by requiring the continuity of the wave function $\chi^{J^\pi T}_\nu(r)$ and of its first derivative at the matching radius $r_0$.  The matrix elements of the scattering matrix can then be used to calculate  cross sections and other reaction observables.  For bound-state calculations $\kappa_\nu$ depends on the studied binding energy. Therefore,  the determination of the bound-state energy  and asymptotic normalization constant $C^{J^\pi T}_\nu$ in Eq.~(\ref{bound}) is achieved iteratively starting from an initial guess for the value of the logarithmic derivative of the wave function at the matching radius $r_0$.

In the  $R$-matrix method on Lagrange mesh, the square-integrable functions chosen to expand the wave function in the internal region are Lagrange functions~\cite{Hesse2002184}. This choice greatly simplifies the calculation, particularly in the presence of non local potentials. 
The accuracy of the $R$-matrix method on a Lagrange mesh is such that for a matching radius of $r_0=15$ fm,  $N=25$ mesh points are usually enough to determine a phase shift within the sixth significant digit. The typical matching radius and number of mesh points adopted for the calculations presented in the following section are $r_0=18$ fm and $N=40$.

Finally, the solution of Eq.~(\ref{r-matrix-eq}) is the least expensive of the various computing steps required for a NCSM/RGM calculation, although it can become computationally more involved as the number of coupled channels grows. 

\section{Applications}
\label{applications}
\subsection{Parity-inversion of the $^{11}$Be ground state}
\begin{center}
\begin{table}[t]
\caption{Mean values of the relative kinetic and potential energies and of the internal $^{10}$Be energy in the $^{11}$Be $1/2^+$ ground state. NCSM/RGM calculation as described in the text.}\label{QuaglioniS_tab:2}
\centering
\begin{tabular}{lcccc}
\br\noalign{\smallskip}
NCSM/RGM & $\langle T_{\rm rel} \rangle$  (MeV)& $\langle W\rangle$ (MeV) & $E[^{10}{\rm Be(g.s.,ex.)}]$ (MeV)& $E_{\rm tot}$ (MeV)\\
\noalign{\smallskip}\mr\noalign{\smallskip}
Model Space & $16.65$  & $-15.02$   & $-56.66$ & $-55.03$ \\
Full                  & $\;\,6.56$ & $\;\,-7.39$ & $-57.02$ & $-57.85$\\
\noalign{\smallskip}\br
\end{tabular}
\end{table}
\end{center}
Developing a comprehensive description of nuclear properties requires understanding exotic nuclei, loosely bound system where the neutron-to-proton rations are larger or smaller than those naturally occurring on earth. Among light drip-line nuclei, $^{11}$Be provides a convenient test of several important properties of neutron rich nuclei. The parity-inverted ground state of $^{11}$Be is one of the best examples of disappearance of the $N=8$ magic number with increasing neutron-to-proton ratio. Contrary to the shell model prediction of $\tfrac12^-$~\cite{PhysRevLett.4.469}, the observed ground-state spin-parity of $^{11}$Be is $\tfrac 12^+$.  
In 2005 Forss\'en {\em et al}.\ published large-scale {\em ab initio} NCSM calculations with several accurate $NN$ potentials of the $^{11}$Be low-lying spectrum~\cite{PhysRevC.71.044312}. Despite the large model space adopted, they were not able to explain the g.s.\ parity inversion. 
This result was partly attributed to the size of the HO basis, which was not large enough to reproduce the correct asymptotic of the $n$-$^{10}$Be component of the 11-body wave function. At the same time the calculations performed with the INOY (inside non-local outside Yukawa) $NN$ potential of Doleschall {\em et al}.~\cite{PhysRevC.69.054001} suggested that the use of a realistic $NNN$ force in a large NCSM basis might correct this discrepancy with experiment. 

The correct asymptotic behavior of the $n$-$^{10}$Be wave function within $^{11}$Be can be reproduced when working within the {\em ab initio} NCSM/RGM approach. In particular, we performed $N_{\rm max}=6$ coupled-channel calculations~\cite{PhysRevLett.101.092501,PhysRevC.79.044606} based on $n$-$^{10}$Be channel states with four target eigenstates: ground, $2^+_1$, $2^+_2$, and $1^+_1$ excited states. To facilitate a direct comparison with the earlier NCSM results, we used the same CD-Bonn $NN$ interaction and HO frequency $\hbar\Omega=13$ MeV, as in Ref.~\cite{PhysRevC.71.044312}. Within this model space, we found that the NCSM and NCSM/RGM energies of the $\tfrac12^-$ states are in rough agreement (-57.51 and -57.59 MeV, respectively), whereas they differ by  a dramatic $\sim3.5$ MeV in the case of the $\tfrac12^+$ state, where the $N_{\rm max}=6$ energy is -54.39 MeV in the NCSM and -57.85 MeV in the NCSM/RGM.   As a result, the $\tfrac 12^-$ and $\tfrac 12^+$ NCSM/RGM states are both bound (by 0.42 and 0.68 MeV, respectively) and the $\tfrac 12^+$ state is the g.s.\ of $^{11}$Be. 

To understand the reason of such a striking difference between the NCSM and NCSM/RGM results for the $\tfrac12^+$ energy, we evaluated mean values of the relative kinetic and potential energies as well as the mean value of the $^{10}$Be energy, and compared them to those obtained by restricting all the integration kernels to the HO model space [{\em i.e.} by replacing the delta function of Eq.~(\ref{N-kernel-2}) with its representation in the HO model space]. In this latter case, as in the NCSM, one loses the correct asymptotic behavior of the $n$-$^{10}$Be wave function. Indeed, it can be seen form Table~\ref{QuaglioniS_tab:2} that the model-space-restricted calculation is similar (although not identical) to the standard NCSM calculation. In addition, due to the re-scaling of the relative wave function in the internal region when the Whittaker tail is recovered, in the full NCSM/RGM calculation we find that  both average kinetic and potential energies are smaller in absolute value than those obtained within the HO model space, and this difference is larger for the relative kinetic energy. This is the origin of the dramatic decrease in energy of the $\frac12^+$ state, which makes it bound and even leads to a g.s.\ parity inversion. Although the present calculations are not sufficient to exclude a role of $NNN$ force in the inversion mechanism, it is clear that an accurate understanding of loosely-bound systems can be achieved only within a dynamic approach that encompasses the continuum.

\subsection{$^4$He$(N,N)^4$He scattering}
The simplest system to be described in terms of binary-cluster basis states of the type described in Eq.~(\ref{basis}) is the scattering of nucleons on $^4$He targets. Here,  energy arguments suggest that already channel states formed by a nucleon in relative motion with respect to an $^4$He nucleus in its g.s.\ should provide a very good description of this process up to fairly high energies. Indeed, the $^4$He nucleus is tightly bound and its first excited state is more than $20$ MeV above its ground state. At the same time, well-determined scattering amplitudes from R-matrix fits make $n$- and
$p$-$^4$He scattering calculations ideal benchmarks for our {\em ab initio} reaction approach. 

We performed nucleon-$^4$He calculations with the SRG-N$^3$LO $NN$ potential with $\Lambda=2.02$ fm$^{-1}$ in a $N_{\rm max}=17$ NCSM/RGM model space spanned by $N$-$^4$He(g.s.) and $N$-$^4$He$^* (0_2^+)$ channel states. At $N_{\rm max}=17$ ($16$, for the positive parity states) convergence of the HO expansion for the localized parts of the NCSM/RGM integration kernels and for the $^4$He ground- and the first-excited $0^+ 0$ states has been fully reached with this soft $NN$ interaction. 
\begin{figure}[t]
\begin{minipage}{17pc}
\includegraphics[width=17pc]{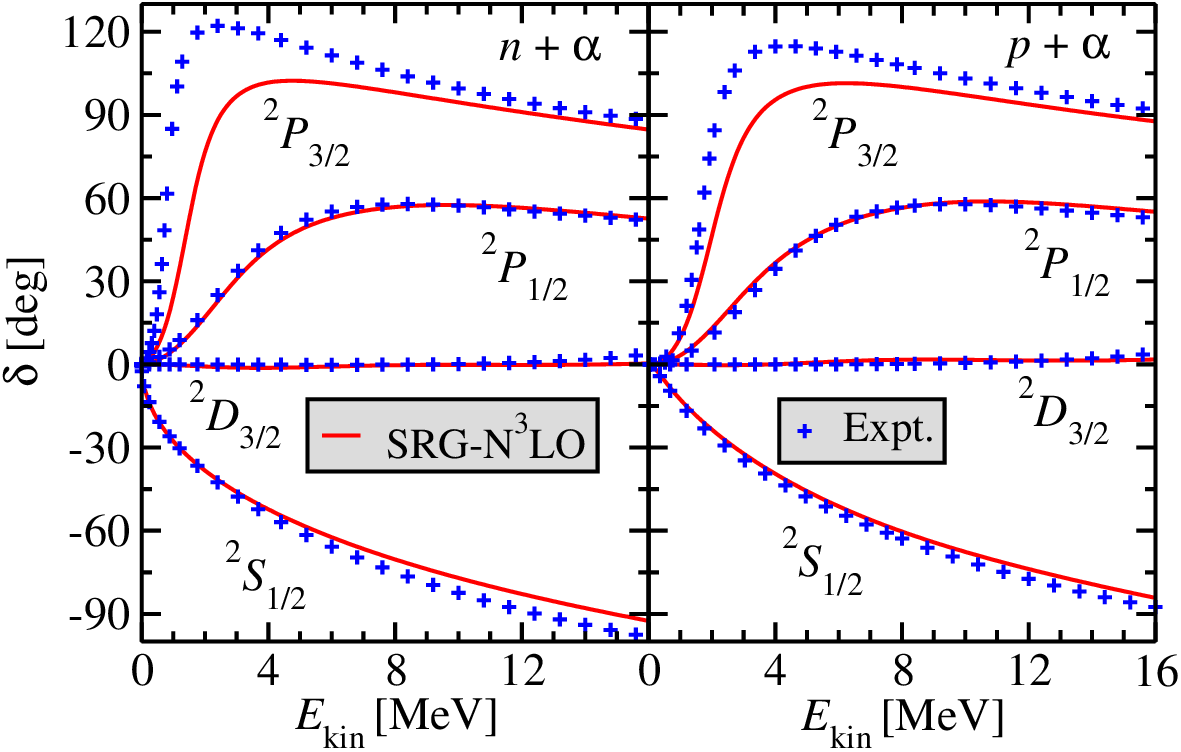}
\caption{\label{fig:N4He_phase}Calculated $n-^4$He (left panels) and $p-^4$He (right panels) compared to the R-matrix analysis of experimental data~\cite{HalePriv}. The NCSM/RGM calculations that included the $^4$He g.s.\ and the $0^+ 0$ excited state were done using the SRG-N$^3$LO $NN$ potential of Ref.~\cite{PhysRevC.77.064003} with a cutoff of 2.02 fm$^{-1}$. The HO frequency $\hbar\Omega=20$ MeV and $N_{\rm max}=17$ basis space were employed.}
\end{minipage}\hspace{3pc}%
\begin{minipage}{17pc}
\includegraphics[width=17pc]{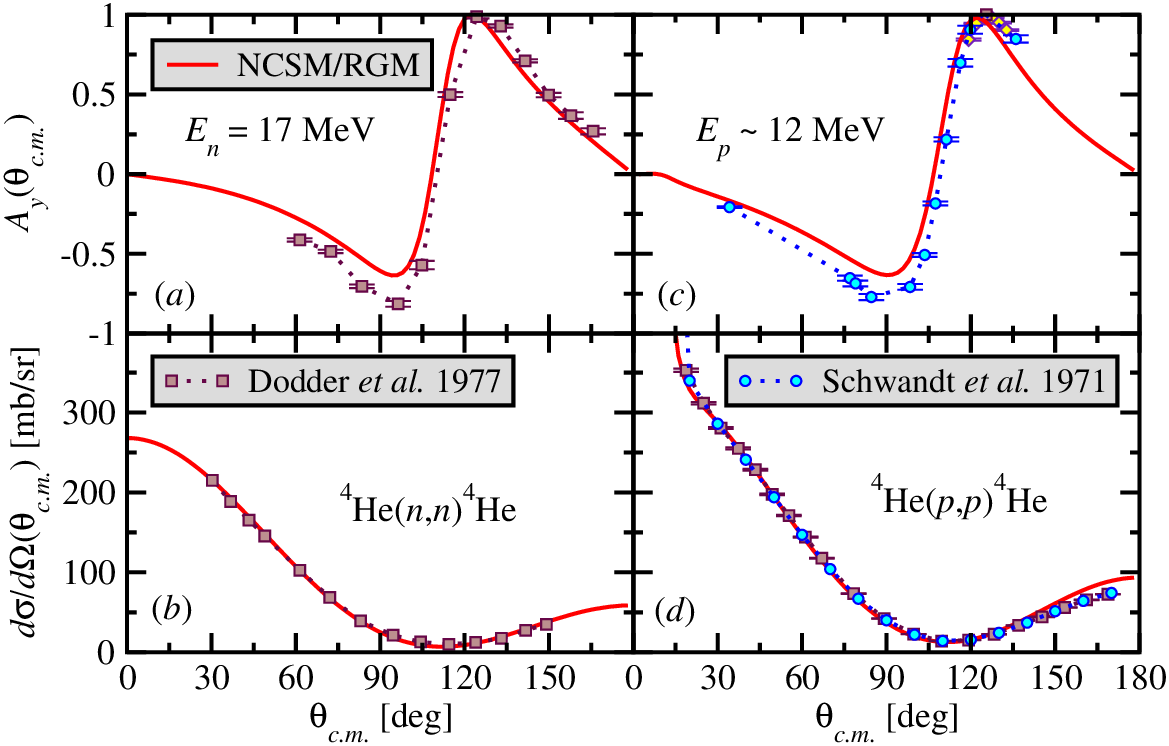}
\caption{\label{cs_Ay}Calculated analyzing power (top panels) and  differential cross section (bottom panels) for $n-^4$He (left panels) at a neutron laboratory energy of $E_n = 17$ MeV and $p-^4$He (right panels) at a proton laboratory energy of $E_p = 12$ MeV compared to experimental data from Schwandt {\em et al}.~\cite{Schwandt}, and  Dodder {\em et al}.~\cite{Dodder}. The NCSM/RGM results as in the caption of Fig.~\ref{fig:N4He_phase}.}
\end{minipage} 
\end{figure}

As expected,  the agreement (shown in Fig.~\ref{fig:N4He_phase}) of our calculated $n$-$^4$He and $p$-$^4$He phase shifts with those obtained from an accurate $R$-matrix analysis of the data is quite reasonable, particularly for c.m.\ energies above $\sim8$ MeV. Correspondingly, in that energy range we can reproduce fairly well also cross-section and polarization data. As an example,  Fig.~\ref{cs_Ay} compares NCSM/RGM $n-$ and $p-^4$He results to the experimental data sets of Dodder {\em et al}.~\cite{Dodder} and Schwandt {\em et al}.~\cite{Schwandt} at $E_n=17$ and $E_p=12$ MeV nucleon laboratory energies, respectively.  The discrepancies observed in the $^2P_{3/2}$ channel, where the calculated resonance is positioned at higher energy and the phase shifts are underestimated with respect to experiment, are largely due to a reduction in spin-orbit strength caused by the omission  in our calculation of the $NNN$ interaction (chiral and SRG-induced). Efforts to include the $NNN$ force into the NCSM/RGM formalism are currently under way. More details on these calculations can be found in Ref.~\cite{PhysRevC.82.034609}. 

\subsection{Accurate evaluation of the $^3$H$(n,n)^3$H cross section}
\begin{figure}[t]
\begin{minipage}{17pc}
\includegraphics[width=17pc]{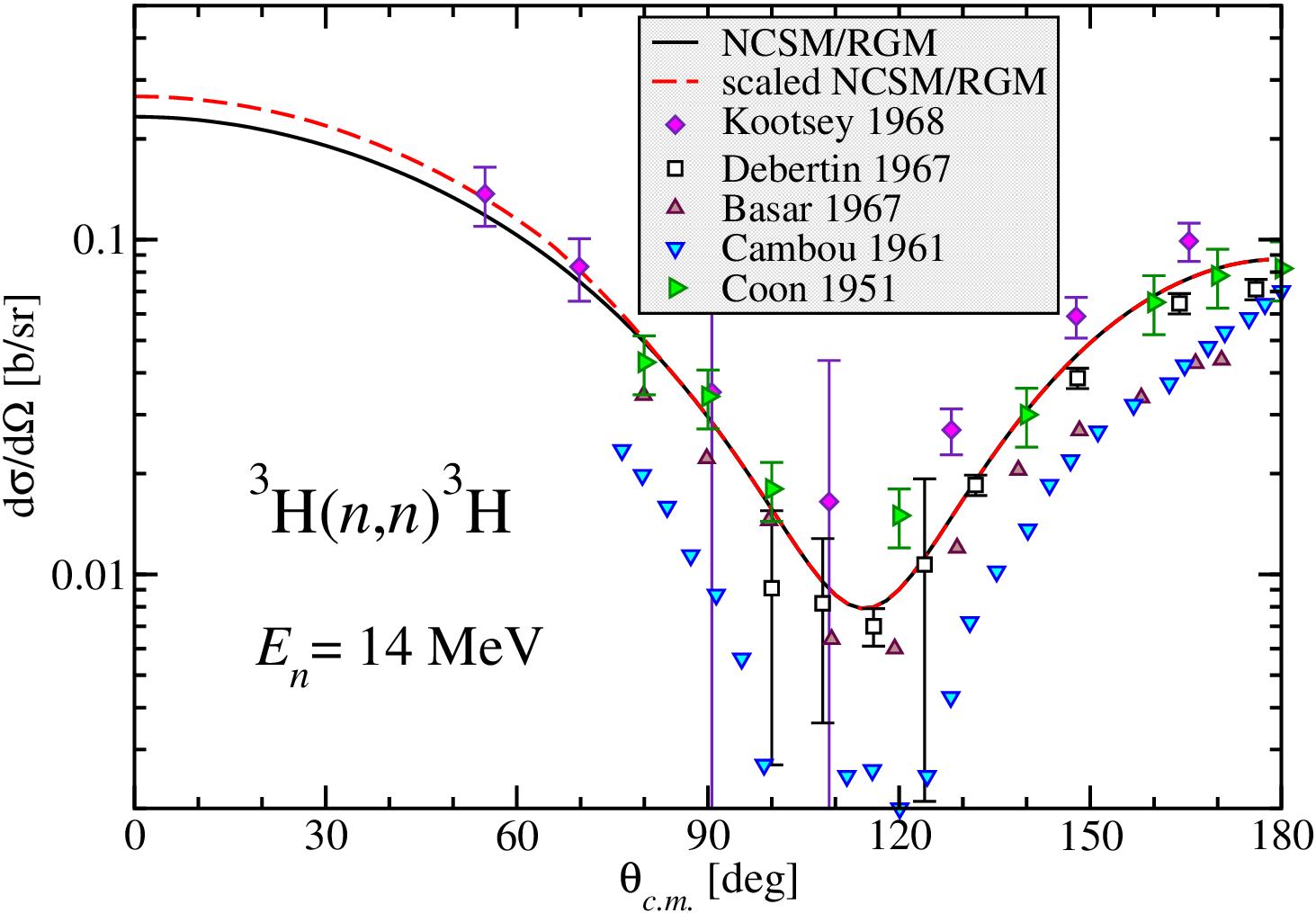}
\caption{\label{fig:n3H}Neutron-$^3$H elastic differential cross section at 14 MeV. The NCSM/RGM (solid black curve) and scaled NCSM/RGM (dashed red curve) calculations described in the text are compared to available beam-target data~[all].}
\end{minipage}\hspace{3pc}%
\begin{minipage}{17pc}
\includegraphics[width=17pc]{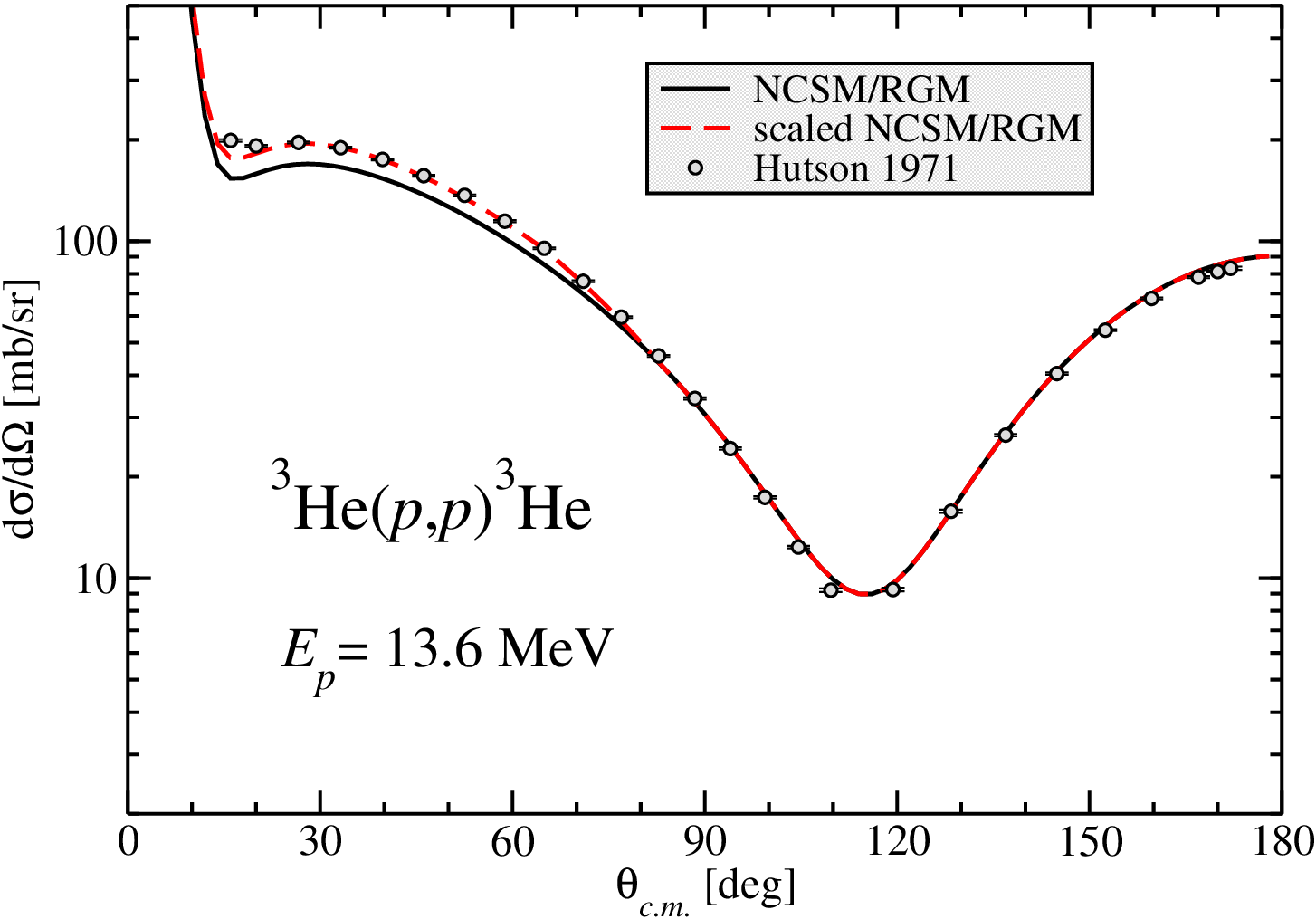}
\caption{\label{fig:p3He}Proton-$^3$He elastic differential cross section at 13.6 MeV. The NCSM/RGM (solid black line) and scaled NCSM/RGM (dashed red curve) calculations described in the text are compared to the experimental data of Hutson {\em et al.}~\cite{Hutson71}.}
\end{minipage} 
\end{figure}

The elastic $n-^3$H cross section induced by $14.1$ MeV neutrons is important for understanding how the fuel is assembled in an inertial confinement implosion based on deuterium-tritium fuel such as those occurring at NIF~\cite{NIF}. To reliably infer a fuel areal density from the yield ratio between scattered and primary $14.1$ MeV neutrons, this process needs to be known within about $5\%$ accuracy. However, as shown in Fig.~\ref{fig:n3H}, the available data from beam-target experiments draw a rather uncertain picture for the $^3$H$(n,n)^3$H differential cross section at this energy. In contrast, the elastic differential cross section for the mirror reaction, $p$-$^3$He, was measured with high accuracy at $E_p$=13.6 MeV~\cite{Hutson71}.  

In an attempt to provide an improved evaluation for the $14.1$ Mev $n$-$^3$H elastic cross section, we performed NCSM/RGM calculations for both this and the mirror ($p$-$^3$H) process, using $n$-$^3$H(g.s.) and $p$-$^3$He(g.s.) channel states, respectively, in an HO model space with $N_{\rm max}=17$ and $\hbar\Omega=20$ MeV. As for the $N-^4$He case described in the previous section, we adopted the SRG-N$^3$LO $NN$ potential with $\Lambda=2.02$ fm$^{-1}$. The  results of these calculations correspond to the black solid curve of Figs.~\ref{fig:n3H} and~\ref{fig:p3He}. The agreement with the experimental $^3$He$(p,p)^3$He differential cross section of  Hutson {\em et al.}~\cite{Hutson71} is very good at backward angles. At forward angles, the data are underestimated by up to 15\%. Such inaccuracy in the NCSM/RGM calculations is largely due to the omission of  channel states with the three-nucleon system in a breakup state, which are obviously important at the laboratory energies considered here. While the inclusion of such channels was out of reach of the present calculation,  we used the $p$-$^3$He data to quantify and correct for this inaccuracy by deducing a smooth scaling factor and applying it to the calculated $n$-$^3$H differential cross section. The result of this procedure, estimated to be accurate to $\sim5\%$, are shown by red dashed curves (scaled NCSM/RGM) in Figs.~\ref{fig:n3H} and~\ref{fig:p3He}. The $n$-$^3$H differential cross section at $E_n$=14 MeV inferred in this way, compares well with that obtained by means of an R-matrix analysis (also relying on the $p$-$^3$He data)~\cite{Hale90} and with a new set of measurements obtained in a deuterium-tritium inertial confinement implosion at the OMEGA laser~\cite{PhysRevLett.107.122502}. The integrated elastic cross section at 14 MeV obtained from the scaled NCSM/RGM calculation is 0.94 barn.  Finally, we note that there is currently no {\em ab initio} theory capable of precisely describing the $n$-$^3$H reaction above the $^3$H breakup threshold, where, in addition to $n$-$^3$H, one should include also three- ($n$-$n$-$d$) and  four-body ($n$-$n$-$n$-$p$) final states. Efforts to extend the NCSM/RGM approach to include three-cluster channel states are currently under way. 

\subsection{The $^7$Be$(p,\gamma)^8$B radiative capture}
The $^7$Be$(p,\gamma)^8$B radiative capture is the final step in the nucleosynthetic chain leading to $^8$B and one of the main inputs of the standard model of solar neutrinos. Recently, we have performed the first {\em ab initio} many-body calculation~\cite{Navratil2011379}, 
of this reaction starting from the SRG-N$^3$LO $NN$ interaction with $\Lambda=1.86$ fm$^{-1}$. Using $p$-$^7$Be channel states including the five lowest $N_{\rm max}=10$ eigenstates of $^7$Be (the $\tfrac32^-$ ground and the $\tfrac12^-$,$\tfrac72^-$, and first and second $\tfrac52^-$ excited states), we solved Eq.~(\ref{RGMeq}) first with bound-state boundary conditions to find the bound state of $^8$B, and then with scattering boundary conditions to find the $p$-$^7$Be scattering wave functions. Former and latter wave functions were later used to calculate the capture cross section, which, at solar energies, is dominated by non-resonant $E1$ transitions from $p$-$^7$Be $S$- and $D$-waves into the weakly-bound ground state of $^8$B.  All stages of the calculation were based on the same HO frequency of $\hbar\Omega=18$ MeV, which minimizes the g.s.\ energy of $^7$Be. The largest model space achievable for the present calculation within the full NCSM basis is $N_{\rm max}=10$. At this basis size,  the $^7$Be g.s. energy is very close to convergence as indicated by a fairly flat frequency dependence in the range $16\le\hbar\Omega\le20$ MeV, and the vicinity to the $N_{\rm max}=12$ result obtained within the importance-truncated NCSM~\cite{PhysRevLett.99.092501,PhysRevC.79.064324}.  The choice of $\Lambda=1.86$  fm$^{-1}$ in the SRG evolution of the N$^3$LO $NN$ interaction leads to a single $2^+$ bound state for $^8$B with a separation energy of 136 keV quite close to the observed one (137 keV). This is very important for the description of the low-energy behavior of the  $^7$Be$(p,\gamma)^8$B astrophysical S-factor, known as $S_{17}$. We note that the $NNN$ interaction induced by the SRG evolution of the $NN$ potential is repulsive in the $\Lambda$-range $\sim1.8$-$2.1$ fm$^{-1}$, and, in very light nuclei, its contributions are canceled to a good extent by those of the initial attractive chiral $NNN$ force (which is also SRG evolved)~\cite{PhysRevLett.103.082501,PhysRevC.83.034301}.   
\begin{figure}[t]
\begin{minipage}{17pc}
\includegraphics[width=17pc]{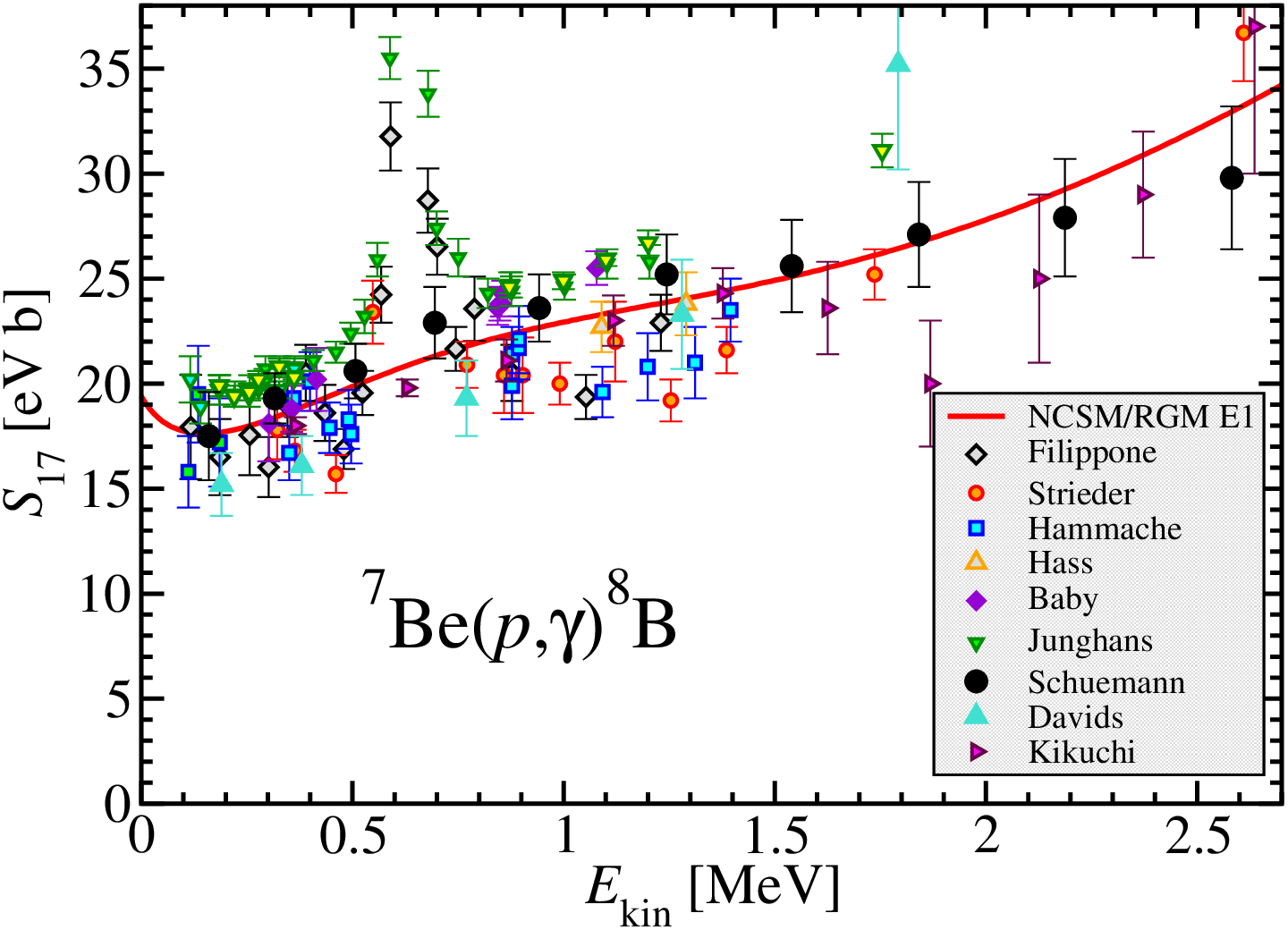}
\caption{\label{fig:p7Be1}Calculated $^7$Be$(p,\gamma)^8$B S-factor as a function of the energy in the center of mass compared to data. Only $E1$ transition were considered in the calculation.}
\end{minipage}\hspace{3pc}%
\begin{minipage}{17pc}
\includegraphics[width=17pc]{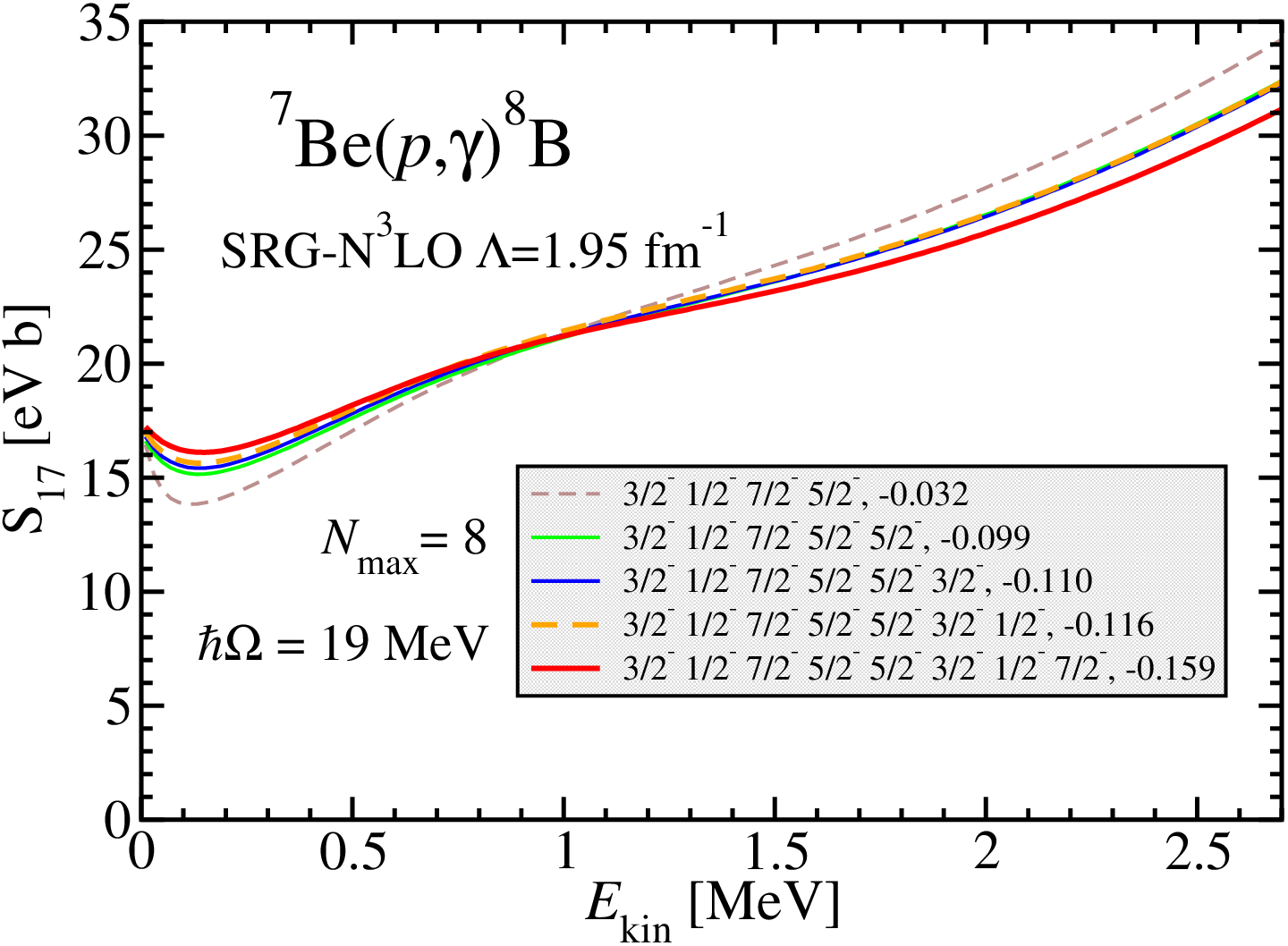}
\caption{\label{fig:p7Be2}Convergence of the $^7$Be$(p,\gamma)^8$B S-factor as a function of the number of $^7$Be eigenstates included in the calculation (shown in the legend together with the corresponding separation energy). }
\end{minipage} 
\end{figure}

The resulting $S_{17}$ astrophysical factor is compared to several experimental data sets in Figure~\ref{fig:p7Be1}. Energy dependence and absolute magnitude follow closely the trend of the indirect Coulomb breakup measurements of Sh\"umann {et al}.~\cite{Schuemann1,Schuemann2}, while somewhat underestimating the direct data of Junghans {\em et al}.~\cite{Junghans}. The resonance, particularly evident in these and Filippone's data, is due to the $M1$ capture, which does not contribute to a theoretical calculation outside of the narrow $^8$B $1^+$ resonance and is negligible at astrophysical energies~\cite{Adelberger1,Adelberger2}. The $M1$ operator, for which any dependence upon two-body currents needs to be included explicitly, poses more uncertainties than the Siegert's $E1$ operator. At the same time, the treatment of this operator within the NCSM/RGM approach is slightly complicated by the additional contributions coming from the core ($^7$Be) part of the wave function. Nevertheless, we plan to calculate its contribution in the future. 

The convergence of our results with respect to the size of the HO model space was assessed by means of calculations up to $N_{\rm max}=12$ within the importance-truncation NCSM scheme~\cite{PhysRevLett.99.092501,PhysRevC.79.064324}with (due to computational limitations) only the first three eigenstates of $^7$Be. The $N_{\rm max}=10$ and $12$ S-factors are very close. As for the convergence in the number of $^7$Be states, we explored it by means of calculations including up to 8 $^7$Be eigenstates in a $N_{\rm max}=8$ basis (larger $N_{\rm max}$ values are currently out of reach with more then five $^7$Be states).  This last set of calculations is presented in Fig.~\ref{fig:p7Be2}, from which it appears that, apart from the two $\tfrac52^-$ states, the only other state to have a significant impact on the $S_{17}$ is the second $\frac72^-$, the inclusion of which affects the separation energy and contributes somewhat to the flattening of the $S$-factor around $1.5$ MeV.  We note that for these last set of calculations we used SRG-N$^3$LO interactions obtained with different $\Lambda$  values with the intent to math closely the experimental separation energy in each of the largest model spaces. Based on this analysis, we conclude that the use of an $N_{\rm max}=10$ HO model space is justified and the limitation to five $^7$Be eigenstates is quite reasonable. Finally, our calculated $S_{17}(0)=19.4(7)$ MeV b is on the lower side, but consistent with the latest evaluation $20.8\pm0.7$(expt)$\pm1.4$(theory)~\cite{Adelberger2}. 

\subsection{The $^3$H$(d,n)^4$He and $^3$He$(d,p)^4$He fusion reactions}
\begin{figure}[t]
\begin{minipage}{17pc}
\includegraphics[width=17pc]{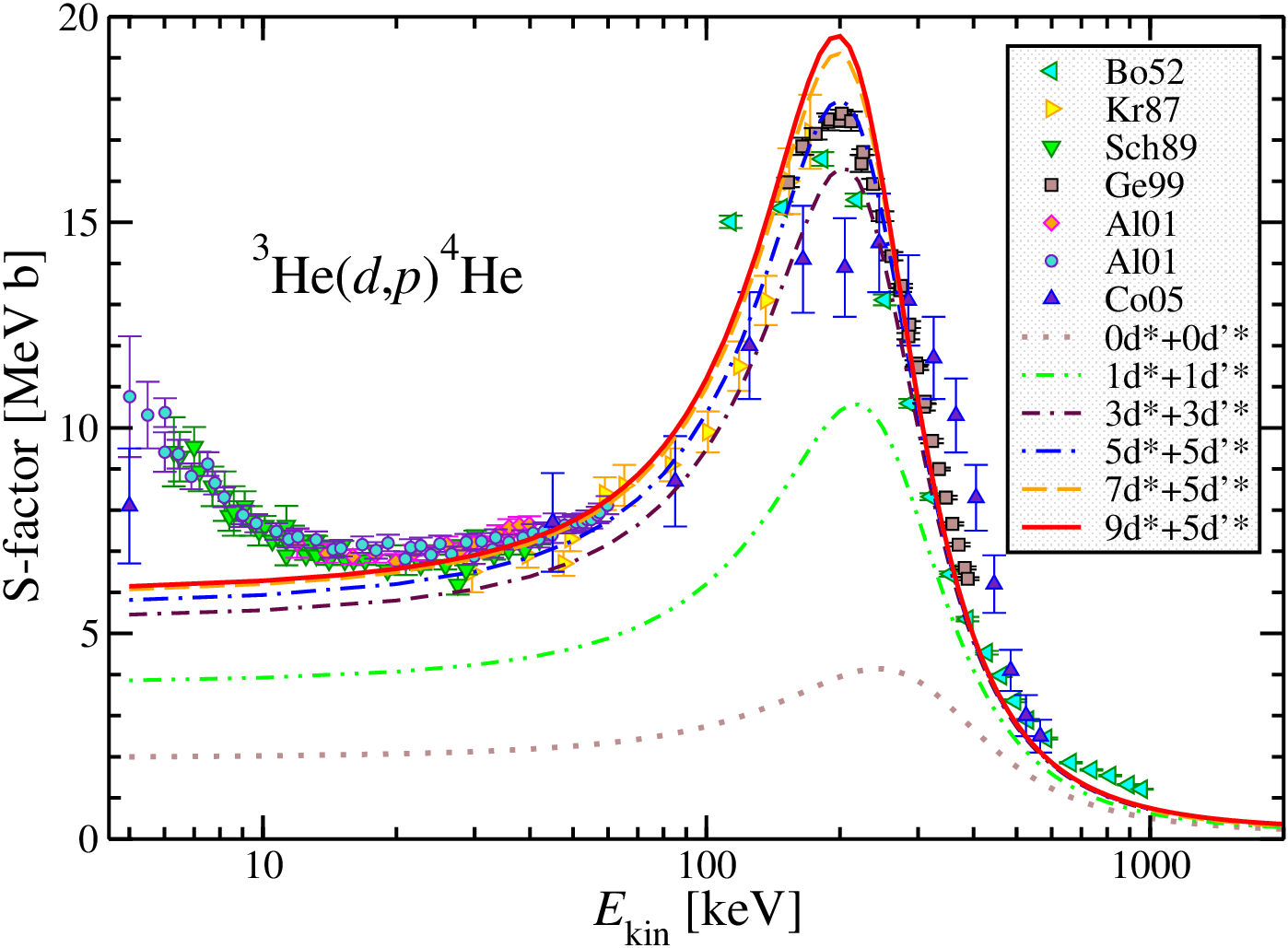}
\caption{\label{fig:d3He}Calculated S-factor of the $^3$He$(d,p)^4$He reaction compared to experimental data. Convergence with the number of deuterium pseudostates in the $^3S_1$-$^3D_1$ ($d^*$) and $^3D_2$ ($d^{\prime *}$) channels. }
\end{minipage}\hspace{3pc}%
\begin{minipage}{17pc}
\includegraphics[width=17pc]{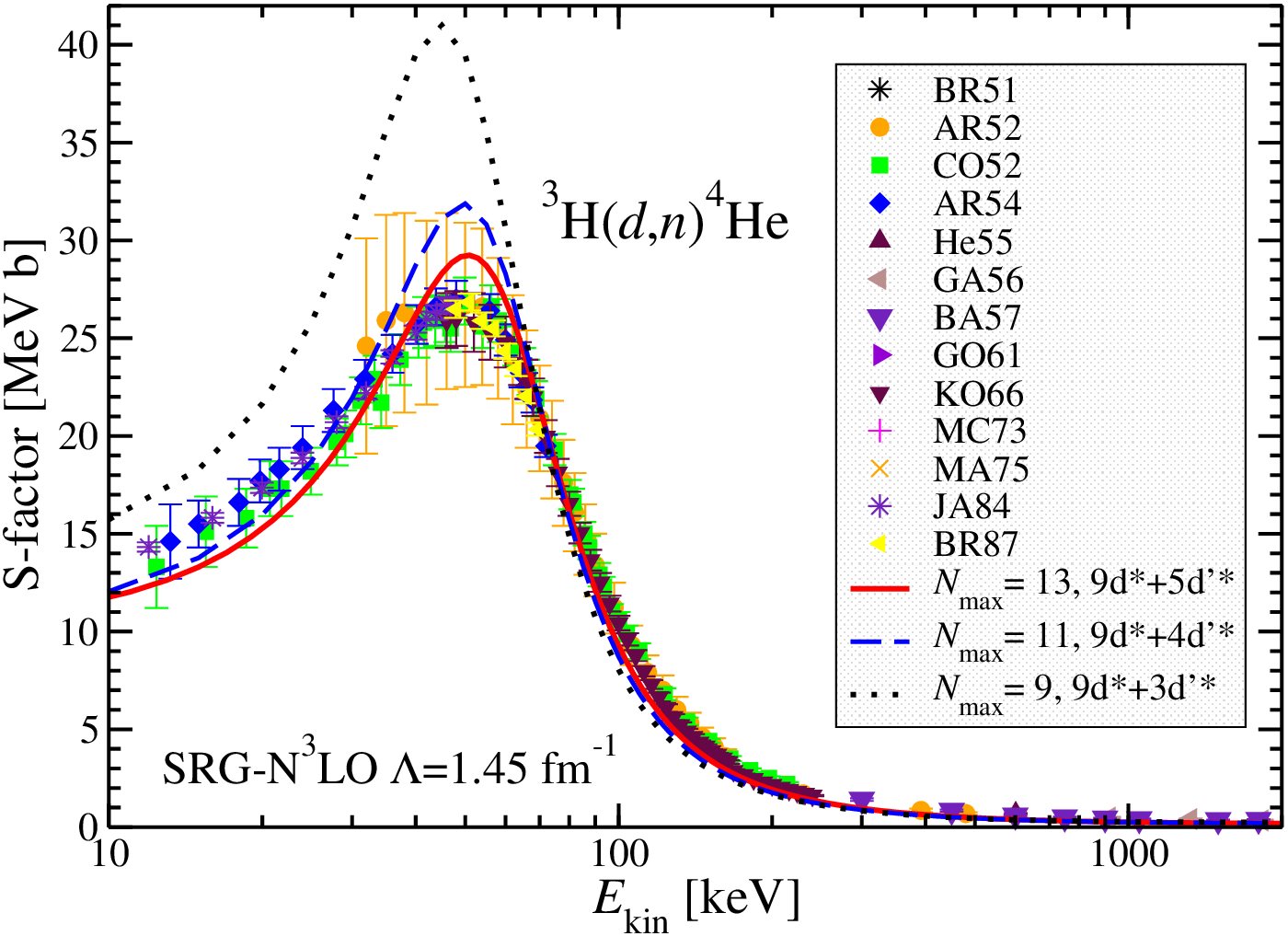}
\caption{\label{fig:d3H}Calculated $^3$H$(d,n)^4$He S-factor compared to experimental data. Convergence with $N_{\rm max}$ obtained for the SRG-N$^3$LO $NN$ potential with $\Lambda=1.45$ fm$^{-1}$ at $\hbar\Omega=14$ MeV.}
\end{minipage} 
\end{figure}
The $^3$H$(d,n)^4$He and $^3$He$(d,p)^4$He fusion reactions have important implications first and foremost for fusion energy generation, but also for nuclear astrophysics, and atomic physics. Indeed, the deuterium-tritium fusion is the easiest reaction to achieve on earth and is pursued by research facilities directed at reaching fusion power by either inertial ({\em e.g.}, NIF) or magnetic ({\em e.g.}, ITER) confinement. Both $^3$H$(d,n)^4$He and $^3$He$(d,p)^4$He  affect the predictions of Big Bang nuclosynthesis for light-nucleus abundances. In addition, the deuterium-$^3$He fusion is also an object of interest for atomic physics, due to the substantial electron-screening effects presented by this reaction. 

In the following we present the first {\em ab initio} many-body calculations~\cite{PhysRevLett.108.042503} of these reactions starting from the SRG-N$^3$LO $NN$ interaction with $\Lambda=1.5$ fm$^{-1}$, for which we reproduce the experimental $Q$-value of both reactions within $1\%$. We adopted HO model spaces up to $N_{\rm max}=13$ with a frequency of $\hbar\Omega=14$ MeV. The channel basis includes $n$-$^4$He ($p$-$^4$He), $d$-$^3$H ($d$-$^3$He), $d^*$-$^3$H ($d^*$-$^3$He) and $d^{\prime*}$-$^3$H ($d^{\prime*}$-$^3$He) binary cluster states, where $d^*$ and $d^{\prime*}$ denote $^3S_1$-$^3D_1$ and $^3D_2$ deuterium excited pseudostates, respectively, and the $^3$H ($^3$He) and $^4$He nuclei are in their ground state. 

Figure~\ref{fig:d3He} presents the results obtained for the $^3$He$(d,p)^4$He S-factor. The deuteron deformation and its virtual breakup, approximated by means of $d$ pseudosates, play a crucial role. The S-factor increases dramatically with the number of pseudostates until convergence is reached for $9d^*+5d^{\prime*}$. The dependence upon the HO basis size is illustrated by the $^3$H$(d,n)^4$He results of Fig.~\ref{fig:d3H}. The convergence is satisfactory and we expect that an $N_{\rm max}=15$ calculation, which is currently out of reach, would not yield significantly different results. The experimental position of the $^3$He$(d,p)^4$He S-factor is reproduced within few tens of keV. Correspondingly, we find an overall fair agreement with experiment for this reaction, if we exclude the region at very low energy, where the accelerator data are enhanced by laboratory electron screening. 
The $^3$H$(d,n)^4$He S-factor is not described as well with  $\Lambda=1.5$ fm$^{-1}$. Due to the very low activation energy of this reaction, the S-factor (particularly peak position and height) is extremely sensitive to higher-order effects in the nuclear interaction, such as three-nucleon force (not yet included in the calculation) and missing isospin-breaking effects in the integration kernels (which are obtained in the isospin formalism).  To compensate for these missing higher-order effects in the interaction and reproduce the position of the  $^3$H$(d,n)^4$He S-factor, we performed additional calculations using lower $\Lambda$ values. This led to the theoretical S-factor of Fig.~\ref{fig:d3H} (obtained for $\Lambda=1.45$ fm$^{-1}$), that is in overall better agreement with data, although it presents a slightly narrower and somewhat overestimated peak. This calculation would suggest that some electron-screening enhancement could also be present in the $^3$H$(d,n)^4$He measured S factor below ~10 keV c.m.\ energy. However, these results cannot be considered conclusive until more accurate calculations using a complete nuclear interaction (that includes the three-nucleon force) are performed. Work in this direction is under way. 
\section{Conclusions and Outlook}
\label{conclusions}
\begin{figure}[t]
\begin{minipage}{17pc}
\includegraphics[width=17pc]{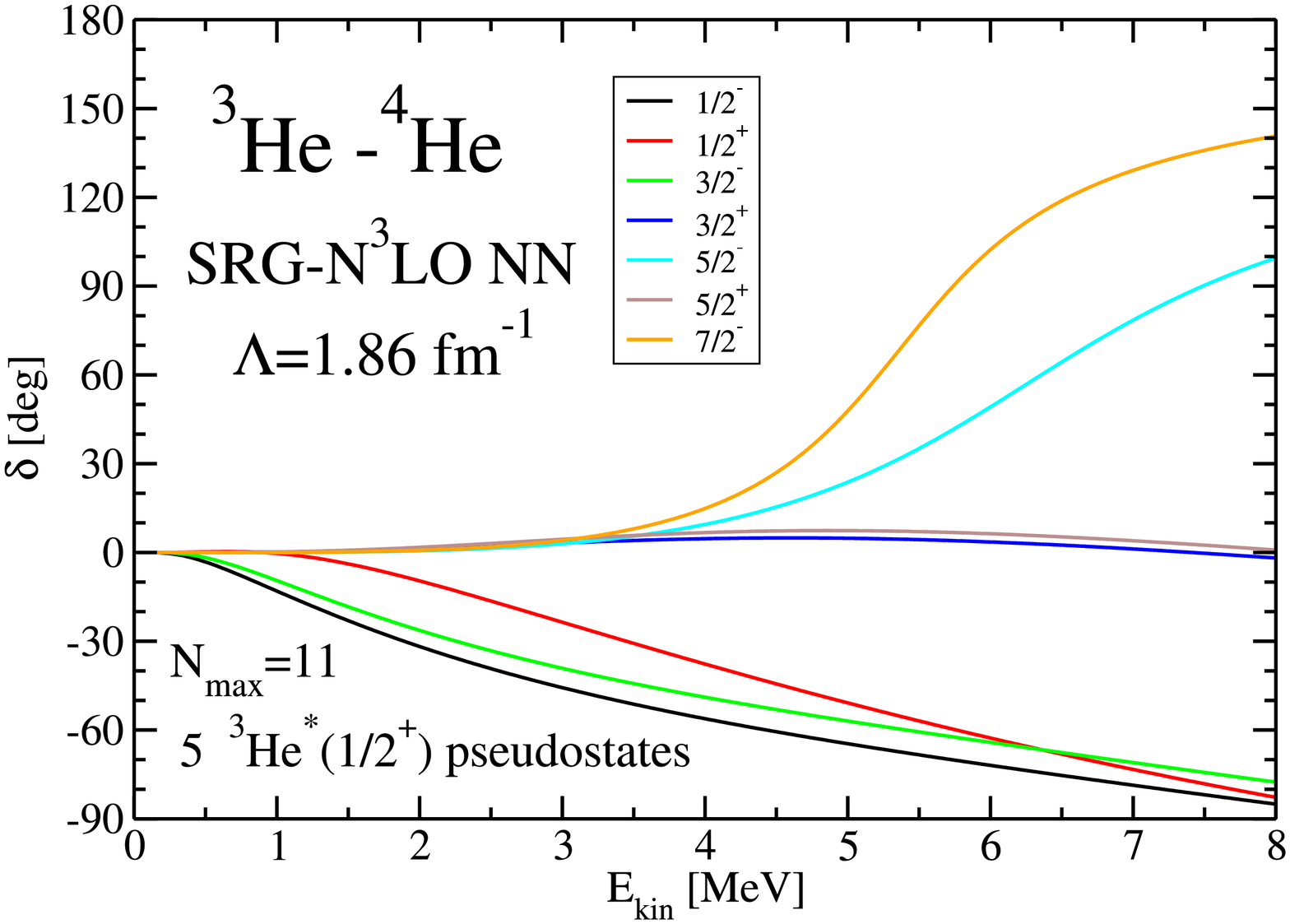}
\caption{\label{fig:He3He4} Preliminary results for the $^3$He-$^4$He scattering phase shifts. The NCSM/RGM calculations, including the g.s. and first four $1/2^+$ pseudostates of $^3$He, were obtained using the SRG-N$^3$LO $NN$ potential with $\Lambda=1.86$ fm$^{-1}$. The HO frequency $\hbar\Omega=18$ MeV and $N_{\rm max}=11$ basis space were employed.  }
\end{minipage}\hspace{3pc}%
\begin{minipage}{17pc}
\rotatebox{-90}{\includegraphics[width=13pc]{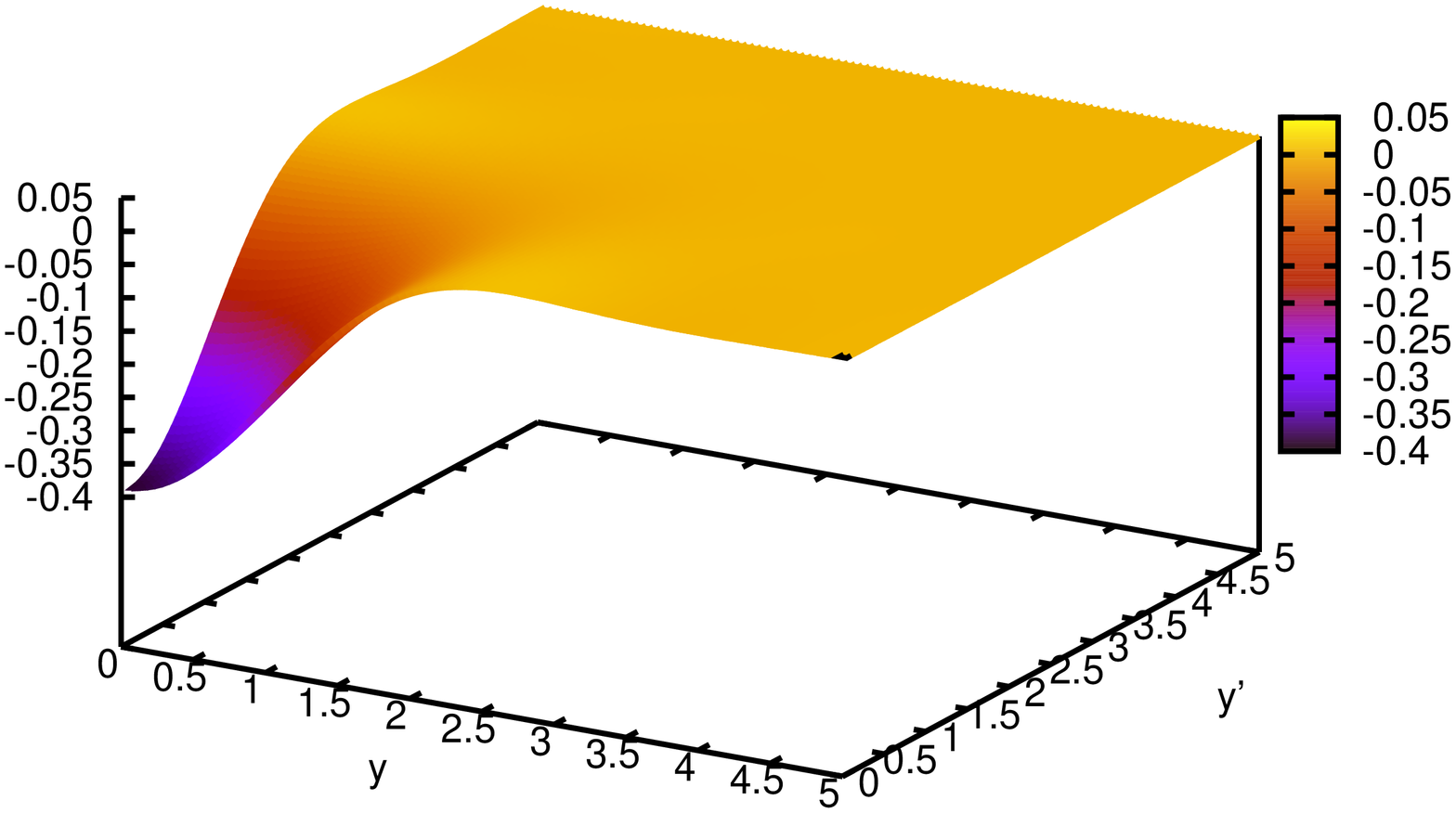}}
\caption{\label{fig:norm} Preliminary results for the exchange part of the $n$+$n$+$^4$He norm kernel (in units of fm$^{-6}$) in the relative channel with the two neutrons (with parallel spins) in $P$-wave motion with respect to each other, and in overall $S$-wave motion with respect to the $^4$He nucleus as a function of their distance $y$ and $y^\prime$ from the target. The two neutrons in the initial and final states are at a fixed distance of $2$ fm.}
\end{minipage} 
\end{figure}
We gave an overview of the NCSM/RGM,  an {\em ab initio} many-body approach capable 
of providing a unified description of structural and reaction properties of light nuclei,
by combining the RGM with the use of realistic interactions, and a microscopic and consistent description of the nucleon clusters, achieved via the {\em ab initio} NCSM. 

Since the publication of the first results~\cite{PhysRevLett.101.092501,PhysRevC.79.044606,PhysRevC.82.034609}, obtained for nucleon-nucleus collisions, the NSCM/RGM has grown into a powerful approach for the description of light-ion fusion reactions. The formalism has been extended to include two-nucleon (deuteron) projectiles~\cite{PhysRevC.83.044609}, as well as complex reactions with both nucleon-nucleus and deuteron-nucleus channels~\cite{PhysRevLett.108.042503}, based on realistic $NN$ interactions. The treatment of three-nucleon (triton and $^3$He) projectiles has also been included in the formalism, and will soon allow the first {\em ab initio} calculation of the $^3$He$(\alpha,\gamma)^7$Be radiative capture. Figure~\ref{fig:He3He4} shows preliminary results of our ongoing effort toward this goal. Further extensions of the approach to include the three-nucleon components of the nuclear interaction and three-cluster channel states are under way. Preliminary results for the exchange part of the $n$-$n$-$^4$He norm kernel are shown in Fig.~\ref{fig:norm}. 

In this contribution, we have revisited the general formalism on which the NCSM/RGM is based, and given the expressions for the terms entering the integrations kernels in the case of a deuteron-nucleus initial and nucleon-nucleus final states, as well as the additional terms appearing in the single-nucleon projectile potential kernel, when the three-nucleon force is included in the Hamiltonian.

Among the applications, we reviewed: the parity inversion of the $^{11}$Be nucleus and nucleon scattering on $^4$He, benchmark calculations that demonstrate the suitability of the NCSM/RGM approach for the description of loosely bound nuclei and low-energy scattering; the evaluation of the $^3$H$(n,n)^3$H differential cross section for 14 MeV laboratory neutrons, an example of how the {\em ab initio} NCSM/RGM can provide the research community with accurate evaluations and uncertainties for less known reactions important for fusion-energy research;  and the first many-body {\em ab initio} calculations for the astrophysically important $^7$Be$(p,\gamma)^8$B radiative capture and the landmark $^3$H$(d,n)^4$He and $^3$He$(d,p)^4$He fusion reactions. These results are very promising and pave the way for achieving improved evaluations of reactions relevant to astrophysics and fusion research, such as the $^3$He$(\alpha,\gamma)^7$Be radiative capture or the $^3$H$+d \rightarrow ^4$He$+n+\gamma$ bremsstrahlung process. 

To conclude, we note that {\em ab initio} NCSM/RGM calculations such as those described in this contributions, and, more importantly, those planned for the future can rapidly grow (with projectile mass, number of projectile/target excited states, and number of channels included) into peta- and even exa-scale computing problems. As an example, the preliminary results of our ongoing investigation of $^3$He-$^4$He scattering presented in Fig.~\ref{fig:He3He4} required runs with up to 64,000 cores on the Oak Ridge National Laboratory (ORNL) Jaguar~\cite{jaguar} supercomputer.   

\ack
Computing support for this work came from the LLNL institutional Computing Grand Challenge program, the J\"ulich supercomputer Centre and Oak Ridge Leadership Computing Facility at ORNL~\cite{jaguar}. Prepared in part by LLNL under Contract DE-AC52-07NA27344.
Support from the U.\ S.\ DOE/SC/NP (Work Proposal No.\ SCW1158), LLNL LDRD grant PLS-09-ERD-020, U.\ S.\ DOE/SC/NP (Work Proposal No.\ SCW0498), the NSERC Grant No. 401945-2011, and from the U.\ S.\ Department of Energy Grant DE-FC02-07ER41457 is acknowledged. This work is supported in part by the Deutsche Forschungsgemeinschaft through contract SFB 634 and by the Helmholtz International Center for FAIR within the framework of the LOEWE program launched by the State of Hesse. W.H. was supported by the
Special Postdoctoral Researchers Program of RIKEN.

\section*{References}

\providecommand{\newblock}{}

\end{document}